\documentclass[prx,aps,amsmath,amssymb,floatfix,twocolumn]{revtex4-1}

\pdfoutput=1
\usepackage{siunitx}
\usepackage{hyperref}
\usepackage{graphicx}
\usepackage[usenames]{color}
\usepackage{bm}

\definecolor{blue}{rgb}{0,0,1}
\definecolor{red}{rgb}{1,0,0}
\newcommand{\bb}[1]{\textcolor{blue}{#1}}
\newcommand{\rr}[1]{\textcolor{red}{#1}}
\newcommand{\new}[1]{\textcolor{black}{#1}}

\def\cal#1{\mathcal{#1}}
\def\eqq#1{Eq.~(\ref{#1})}

\def\eq#1{(\ref{#1})}

\def\f#1{Fig.~\ref{#1}}

\def\s#1{Section~\ref{#1}}

\def\c#1{~\cite{#1}}

\def\beq{\begin{equation}}
\def\eeq{\end{equation}}
\def\bea{\begin{eqnarray}}
\def\eea{\end{eqnarray}}

\begin{document}

\title{Neural units with time-dependent functionality}

\author{Stephen Whitelam}\email{swhitelam@lbl.gov}
\affiliation{Molecular Foundry, Lawrence Berkeley National Laboratory, 1 Cyclotron Road, Berkeley, CA 94720, USA}

\begin{abstract}
We show that the time-resolved dynamics of an underdamped harmonic oscillator can be used to do multifunctional computation, performing distinct computations at distinct times within a single dynamical trajectory. We consider the amplitude of an oscillator whose inputs influence its frequency. The activity of the oscillator at fixed time is a nonmonotonic function of its inputs, and so it can solve problems such as XOR that are not linearly separable. The activity of the oscillator at fixed input is a nonmonotonic function of time, and so it is multifunctional in a temporal sense, able to carry out distinct nonlinear computations at distinct times within the same dynamical trajectory. We show that a single oscillator, observed at different times, can act as all of the elementary logic gates and can perform binary addition, the latter usually implemented in hardware using 5 logic gates. We show that a set of $n$ oscillators, observed at different times, can perform an arbitrary number of analog-to-$n$-bit digital conversions. We also show that oscillators can be trained by gradient descent to perform distinct classification tasks at distinct times. Computing with time-dependent functionality can be done in or out of equilibrium, and suggests a way of reducing the number of parameters or devices required to do nonlinear computations.
\end{abstract}

\maketitle

\section{Introduction} 

Computing is done by physical processes\c{bennett1982thermodynamics,landauer1991information,wolpert2019stochastic}. Classical computing uses the movement of electrons in silicon chips to perform logical operations\c{ceruzzi2003history}; quantum computing uses superposition and entanglement in qubits to process information\c{horowitz2019quantum}; neuromorphic computing mimics the neural and synaptic activities of the human brain\c{schuman2017survey}; echo-state networks use dynamic reservoirs of neural activity to process sequences\c{sun2020review}; analog computing uses the continuous variation of electrical or mechanical signals to solve problems\c{maclennan2007review,ulmann2013analog,csaba2020coupled}; and thermodynamic computing uses the tendency of physical systems to evolve toward thermal equilibrium to do calculations\c{conte2019thermodynamic,hylton2020thermodynamic,aifer2023thermodynamic}. 

Here we show that the explicit time dependence of a physical dynamics can be used to do multifunctional computation, with a single device able to perform multiple distinct calculations in the course of a dynamical trajectory. We consider the dynamics of a continuous-valued underdamped oscillator. Oscillators can be realized in hardware in many ways\c{von1957non,csaba2020coupled,ciliberto2017experiments}, including by springs\c{frenchvibrationswaves}, cantilevers\c{dago2021information,dago2022virtual}, and electrical circuits\c{harris2015digital,mano2017digital,wang2019oim,chou2019analog,melanson2023thermodynamic}. Computing with oscillators is a concept that dates back to the 1950s\c{von1957non,goto1959parametron}. Most examples of oscillator-based computing focus on the phase of the oscillator as a means of carrying information, and aim to find oscillatory ground states in networks of interacting oscillators\c{csaba2020coupled,bonnin2022coupled}. 
\begin{figure*}[t]
   \centering
   \includegraphics[width=0.9\linewidth]{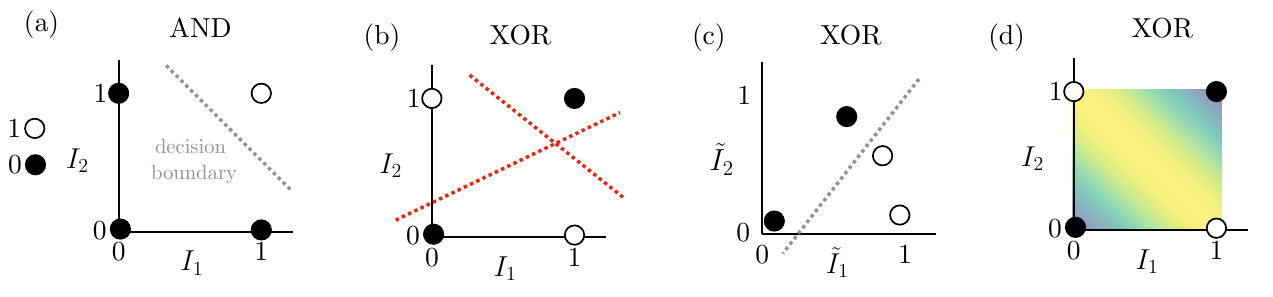} 
   \caption {An illustration of the computational power of neural units with nonmonotonic activations, using binary inputs $I_{1,2}=0,1$. (a) A standard perceptron unit, whose output is a monotonic function of a linear sum of its inputs, can solve AND. (b) However, it cannot solve XOR, which is not linearly separable: neither red line (nor any other straight line) separates the two classes. (c) A multilayer perceptron is required to solve XOR: its hidden layers transform the input data $(I_1,I_2) \to (\tilde{I}_1,\tilde{I}_2)$ until a linear decision boundary suffices. (d) Alternatively, XOR can be solved by a neural unit whose output is a nonmonotonic function of its inputs\c{vodenicarevic2017neural,gidon2020dendritic,noel2021biologically}, here $f(I_1,I_2)=\sin(\pi(I_1+I_2)/2)$. In this paper we build on this observation to show that the dynamics of an underdamped oscillator, whose output is a nonmonotonic function of its inputs {\em and} the elapsed time, can perform distinct nonlinear computations at distinct times.}
   \label{fig_schematic}
\end{figure*}

Here we focus on the time-resolved amplitude of an oscillator whose inputs influence its frequency. The motivation for this choice is twofold. First, because the activity of the oscillator at fixed time is a nonmonotonic function of the input, a single oscillator can solve problems such as XOR that are not linearly separable, making it more expressive than standard artificial neurons. Second, because the activity of the oscillator at fixed input is a nonmonotonic function of time, it is multifunctional in a temporal sense, able to carry out different nonlinear computations at different times within the same dynamical trajectory. 

Units that are a nonmonotonic function of their inputs are more expressive than standard artificial neurons. For example, the standard perceptron unit, which applies a monotonic activation function to a linear function of its inputs, can solve problems that are linearly separable, such as the AND operation\c{rumelhart1986learning}. Consider two binary variables, $I_{1,2}=0,1$. The AND operation returns 1 if both inputs are 1, and returns 0 otherwise. As shown in \f{fig_schematic}(a), a single decision boundary that is a linear function of the two inputs can separate the two output classes, and so a perceptron can solve AND. 

The same is not true of the XOR operation. XOR returns 0 if $I_1$ and $I_2$ are equal, and returns 1 otherwise. As shown graphically in \f{fig_schematic}(b), there is no linear function of the input data for which a monotonic activation function can separate the classes 0 and 1, and so a single perceptron cannot solve XOR. The XOR problem demonstrated the limitations of single-layer perceptrons, leading to the development of multi-layer neural networks and the field of deep learning\c{lecun2015deep,schmidhuber2015deep}. As sketched in \f{fig_schematic}(c), a multilayer perceptron transforms the input data so that a final layer can implement a linear decision boundary that solves the problem. 

Alternatively, XOR can be solved by a single oscillatory unit\c{vodenicarevic2017neural,gidon2020dendritic,noel2021biologically}. The oscillatory function $f(I_1,I_2)=\sin(\pi(I_1+I_2)/2)$, shown as a color map in \f{fig_schematic}(d), satisfies $f(0,0)=f(1,1)=0$ and $f(0,1)=f(1,0)=1$ and so solves XOR. 

\new{Thus a neural unit that is a non-monotonic function of its output is in this sense more powerful than a monotonic activation function. Some neurons in the brain display an output that is a non-monotonic function of their input, enabling them to solve XOR\c{gidon2020dendritic}. Such observations have motivated other authors to show that neural networks built from oscillator units are highly expressive: in Ref.\c{effenberger2022functional}, a recurrent neural network built from damped harmonic oscillators outperformed similar networks built from non-oscillating units, in both learning speed and task performance. While biological neurons and harmonic oscillators are different in detail -- for instance, biological neurons are stochastic, exhibit plasticity, and undergo various biochemical processes that include growth and aging -- an oscillator model can serve as a simplified abstraction of the non-monotonic behavior observed in biological neurons.}

Here we build on these observations by showing that the dynamics of an underdamped oscillator, whose output is a nonmonotonic function of its inputs {\em and} the elapsed time, can perform distinct nonlinear computations at distinct times. In \s{osc} we show that a single oscillator unit can act as {\em all} of the elementary logic gates, depending on the time at which we measure its output, and can do binary arithmetic conventionally done using multiple logic gates. We also show that $n$ oscillators used in parallel, observed at different times, can perform an arbitrary number of analog-to-$n$-bit digital conversions. In \s{grad} we show that oscillators can be trained, by gradient-based adjustment of their input weights, to recognize distinct image types at distinct times. We conclude in \s{conc}. 

A neural unit whose time-resolved activity is the information-carrying element and which oscillates with a frequency influenced by its input could be called an {\em oscillatron}~\footnote{The Oscillatron-1 is neural network whose pattern generators oscillate in synchrony with its feature detectors\c{atkins1991oscillatron}.}, by comparison with {\em perceptron}, the standard monotonic neural unit, and {\em parametron}, an oscillating neural unit whose phase is the information-carrying element\c{goto1959parametron}. Here we will use the terms oscillator, oscillatron, and neuron interchangeably.  A device built from such units could perform multiple computations within a single dynamical trajectory, requiring only a single set of parameters to do multiple tasks.

\section{Dynamical oscillators possess time-dependent functionality}
\label{osc}

Consider a continuous degree of freedom $S(t)$ of unit mass that evolves according to the underdamped Langevin dynamics\c{van1992stochastic} 
\beq
\label{lang}
\ddot{S}=-\gamma \dot{S} -\partial_S U(S)+\eta(t).
\eeq
Here $\gamma$ is a friction coefficient and $U$ is a potential. We will consider the harmonic potential $U(S)= \frac{1}{2} (\omega_0^2+I) S^2$, where $\omega_0$ is an intrinsic frequency and $I$ is an input signal. The input signal could in general be time-dependent, but here we consider a constant input \new{(for instance, constant connections between RLC oscillators can be engineered in circuit boards\c{aifer2023thermodynamic})}. The input therefore influences the spring constant of the harmonic potential. The term $\eta(t)$ in \eq{lang} describes thermal noise, which in general provides an important mechanism for driving dynamical evolution and allowing probabilistic computation\c{aifer2023thermodynamic}. Here we assume the low-noise limit, and so $S(t)$ evolves according to the equation
\beq
\label{unit}
\ddot{S}+\gamma \dot{S} +(\omega_0^2+I) S=0.
\eeq
We impose the initial conditions $S(0)=1$ and $\dot{S}(0)=0$, and so
\beq
\label{sol}
S(t) ={\rm e}^{-\gamma t/2} \left( \cos \Omega t+ \frac{\gamma}{2 \Omega} \sin \Omega t\right),
\eeq
where
\beq
\Omega^2 \equiv \omega_0^2-\gamma^2/4+I
\eeq
is a function of the input $I$ and the oscillator's intrinsic parameters $\omega_0$ and $\gamma$ .

\begin{figure*}
   \centering
   \includegraphics[width=0.9\linewidth]{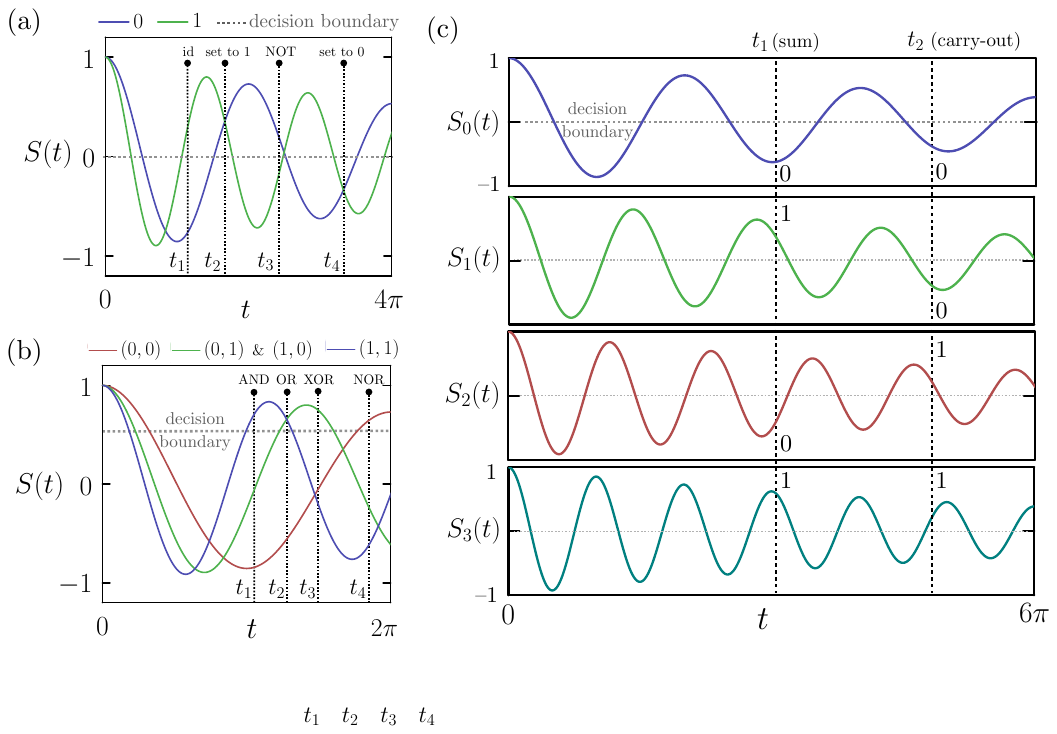} 
   \caption {(a) The time-resolved dynamics \eq{sol} of the oscillatory neural unit \eq{unit}, or oscillatron, evaluated for the two distinct values of a binary input $I=0,1$. Depending on the observation time $t$, the unit can perform any of the elementary one-bit operations. (b) The time-resolved dynamics of the same unit, now evaluated for the three distinct values $I=I_1+I_2=0,1,2$ of the binary inputs $(I_1,I_2)$. Depending on the observation time $t$, the unit can function as any of the elementary logic gates.  (c) A binary full adder realized by the unit \eq{unit} with input $I=A+B+C_{\rm in}$. We show time traces $S_I(t)$, \eqq{sol}, for the four possible values of $I$. The sum $\Sigma$ and carry-out $C_{\rm out}$ are given by the output of the unit at times $t_1$ and $t_2$, respectively. Oscillator parameters: $\omega_0=1, \gamma=1/10$.}
   \label{fig1}
\end{figure*}

We consider the underdamped regime, where $\Omega^2>0$, in which case $S$ oscillates with frequency $\Omega$. $S$ is a damped harmonic oscillator whose input $I$ influences its frequency of oscillation, which in computing terms could be called an oscillatron.  At fixed time $t$ the output of the unit is a nonmonotonic function of $I$, and at fixed $I$ the output of the unit is a nonmonotonic function of $t$. These two properties give this neural unit the ability to solve problems that are not linearly separable, and to solve distinct problems of this nature at distinct times. 

\subsection{Logic gates}

As a first example, consider the binary input $I=0,1$. In \f{fig1}(a) we show the oscillator output \eq{sol} as a function of time $t$ for the two possible values of $I$. With a horizontal decision boundary placed at zero (and values above and below the boundary considered to be states 1 and 0, respectively), the neuron can perform any of the elementary one-bit operations, depending on observation time: identity ($I \to I$) at $t_1$; set (or erase) to one ($I \to 1$) at $t_2$; NOT or invert ($I \to \bar{I}$) at $t_3$; and set to zero ($I \to 0$) at $t_4$. The same operations are also performed at other times in the unit's trajectory. 

Conventional physical models of processes such as bit erasure consider the manipulation of a degree of freedom in an external potential, with work expended\c{landauer1961irreversibility,sagawa2012thermodynamics,zulkowski2014optimal,proesmans2020finite,dago2022dynamics}. Here, by contrast, energy must be input in order to set the initial state of the unit and impose and maintain the input to the neuron (which sets the potential spring constant), but after setting the initial condition no work is performed: computations are done by the system's natural dynamics. Moreover, the same unit can perform different computations when observed at different times.

Now consider two binary input degrees of freedom $I_{1,2}=0,1$, and construct the input $I=I_1+I_2$. \new{(This can be done in hardware using an operational amplifier in summing configuration; the resulting signal sets the spring constant of the oscillator.)}  $I$ can take three values, 0, 1, or 2. In \f{fig1}(b) we show the oscillator output \eq{sol} as a function of time $t$ for the three possible values of $I$. With the horizontal decision boundary placed as shown, the neuron can function as any of the elementary logic gates, depending on observation time. If we observe the unit at time $t_1$ it functions as an AND gate; at time $t_2$, an OR gate; and at time $t_3$, an XOR gate. At different times the same unit can also function as the inverted versions of these gates, NAND, XNOR, and NOR, the latter shown at time $t_4$.

\new{Here we have chosen decision boundaries and observation times by inspection, given a choice of oscillator parameters. In \s{grad} we show how to train oscillator parameters by gradient descent, given a choice of decision boundary and observation times.}

Note that the equilibrium state of the unit is $S(t \to \infty)=0$, regardless of the value of $I$, and so in equilibrium the unit has no computational ability. In general, we can do time-resolved computation in or out of equilibrium, but doing it out of equilibrium means that we do not need to wait for equilibrium to be attained or verify that it has been attained.

\subsection{Binary arithmetic} 

Binary arithmetic on a CPU is done by multiple logic gates\c{harris2015digital,mano2017digital}. For instance, a binary half adder turns binary inputs $A =0,1$ and $B=0,1$ into a sum $\Sigma = A \oplus B$, where $\oplus$ denotes the XOR operation, and a carry $C = A \cdot B$, where $\cdot$ denotes the AND operation. Consequently, the half adder is realized in hardware by two gates, XOR and AND\c{harris2015digital,mano2017digital}. \f{fig1}(b) shows that one oscillator can represent both XOR and AND, at different observation times, indicating that it can function as a half adder: the sum and the carry are the output of the unit at times $t_3$ and $t_1$, respectively. To produce a steady binary signal the output of the unit must be fed to a comparator (to produce a binary signal from an analog one) and a clamp (to produce a time-invariant signal), but the nonlinear computation can be done by a single oscillator. 

This idea can be generalized to more complex operations. A binary full adder turns binary inputs $A$ and $B$ and a carry-in $C_{\rm in}$ into a sum $\Sigma = A \oplus B \oplus C_{\rm in}$ and a carry-out $C_{\rm out} = (A \cdot B) + (B \cdot C_{\rm in}) + (C_{\rm in} \cdot A)$, where $+$ denotes the OR operation\c{harris2015digital,mano2017digital}. The resulting truth table is

\begin{table}[h!]
\centering
\begin{tabular}{ccc|cc}
A & B & $C_{\rm in}$ & $C_{\rm out}$ & $\Sigma$ \\
\hline
0 & 0 & 0 & 0 & 0 \\
0 & 0 & 1 & 0 & 1 \\
0 & 1 & 0 & 0 & 1 \\
0 & 1 & 1 & 1 & 0 \\
1 & 0 & 0 & 0 & 1 \\
1 & 0 & 1 & 1 & 0 \\
1 & 1 & 0 & 1 & 0 \\
1 & 1 & 1 & 1 & 1 \\
\end{tabular}
\caption{Binary full adder: carry-out and sum for the inputs $A$, $B$, and $C_{\rm in}$.}
\end{table}

The full adder is usually realized in hardware by 5 gates, two half adders and an OR gate\c{harris2015digital,mano2017digital}.

In terms of the four possible values of the combination $I=A+B+C_{\rm in}$, the truth table above can be written 

\begin{table}[h!]
\centering
\begin{tabular}{c|cc}
$I$ & $C_{\rm out}$ & $\Sigma$ \\
\hline
0 & 0 & 0 \\
1 & 0 & 1 \\
2 & 1 & 0 \\
3 & 1 & 1 \\
\end{tabular}
\caption{Binary full adder: carry-out and sum for the combination $I=A+B+C_{\rm in}$.}
\end{table}

This truth table can be represented by a single oscillator. In \f{fig1}(c) we show the output $S_I(t)$, \eqq{sol}, of the unit \eq{unit}, for the four possible values of its input $I=A+B+C_{\rm in}$. The output of the unit at time $t_1$ is the required sum, i.e. $\Sigma= S_I(t_1)$, while the output of the unit at time $t_2$ is the required carry-out, i.e. $C_{\rm out}= S_I(t_2)$. Thus one oscillatory unit can perform a computation usually done with 5 logic gates. Again, the output of the oscillator is analog and time-varying, and so must be fed to a comparator and latch in order to provide a steady binary signal, but the computation itself can be done by one device. 

Harmonic oscillators can therefore do multi-bit binary addition: an $n$-bit adder can be built by chaining $n$ full adder devices in series, with the carry-out of each full adder sent to the carry-in of the next full adder. \new{In this case a comparator and latch can function as an encoder, converting the output of one oscillator into a steady digital signal. An operational amplifier in summing configuration can then be used as an encoder, combining that signal with two additional inputs to set the spring constant of the next oscillator in the chain.}

\subsection{Multifunctional analog-to-digital conversion}

In this section we consider a form of analog-to-digital conversion (ADC), with $n$ oscillators used to convert an analog input $A$ into an $n$-bit binary number. The same set of oscillators, observed at $m$ different times, can convert $m$ analog inputs into binary outputs. This section differs from the rest of the paper in that we consider the output of a set of oscillators whose fundamental frequencies are fixed and do {\em not} depend on the input, at observation times that {\em are} a function of the input.

Consider an analog number $A \in [0,A_{\rm max}]$, and form the input 
\beq
\label{aye}
I(A)=\frac{A}{A_{\rm max}} (2^n-1).
\eeq
Here $n$ is the number of binary digits we wish to use to represent $A$. \new{(A linear scaling can be done in hardware using a scaling amplifier or a shift register.)} Table~\ref{tab1} shows the binary representation of the input $I$ sampled at 16 integer values for the case $n=4$, large enough to identify the following pattern: as $I$ increases, the value of each bit $B_i$ oscillates between 0 and 1 with period $2^i$.

\begin{table}[h!]
\centering
\begin{tabular}{c|cccc}
$I$ & \textbf{$B_4$} & \textbf{$B_3$} & \textbf{$B_2$} & \textbf{$B_1$} \\
\hline
0  & \rr{0} & \rr{0} & \rr{0} & \rr{0} \\
1  & \rr{0} & \rr{0} & \rr{0} & \bb{1} \\
2  & \rr{0} & \rr{0} & \bb{1} & 0 \\
3  & \rr{0} & \rr{0} & \bb{1} & 1 \\
4  & \rr{0} & \bb{1} & 0 & 0 \\
5  & \rr{0} & \bb{1} & 0 & 1 \\
6  & \rr{0} & \bb{1} & 1 & 0 \\
7  & \rr{0} & \bb{1} & 1 & 1 \\
8  & \bb{1} & 0 & 0 & 0 \\
9  & \bb{1} & 0 & 0 & 1 \\
10 & \bb{1} & 0 & 1 & 0 \\
11 & \bb{1} & 0 & 1 & 1 \\
12 & \bb{1} & 1 & 0 & 0 \\
13 & \bb{1} & 1 & 0 & 1 \\
14 & \bb{1} & 1 & 1 & 0 \\
15 & \bb{1} & 1 & 1 & 1 \\
\end{tabular}
\caption{\label{tab1} Binary digits $B_i=0,1$ representing an analog input $I$ sampled at integer values. Moving from top to bottom, the period of digit $B_i$ is $2^i$ (the red and blue each indicate a half period), and so can be represented by an oscillator with frequency $\omega_i=2 \pi/2^i$.}
\end{table}

 This pattern indicates that the required bit value can be represented by an oscillator with the appropriate frequency, as shown in \f{fig_adc}. The dashed vertical black line shows the binary representation of an input $14 \leq I < 15$ to be 1110.
 
 \begin{figure}[h]
   \centering
   \includegraphics[width=0.9\linewidth]{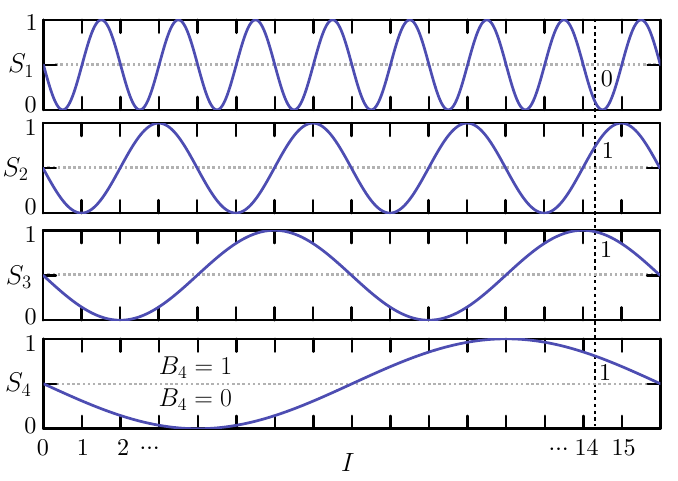} 
   \caption {A graphical representation of Table~\ref{tab1}, showing that the $i^{\rm th}$ binary digit $B_i$ of the analog input $I$ can be represented by an oscillator of period $2^i$, as $B_i =\Theta \left[S_i(t_I)\right]$.}
   \label{fig_adc}
\end{figure}

A set of oscillators of the type \eq{unit} can represent the required set of outputs. Denoting an oscillator's fundamental frequency as $\omega_i$, and setting $\gamma=0$ for simplicity, we impose the initial conditions $S(0)=0$ and $\dot{S}(0)=-1$. Then \eq{sol} is replaced by
\beq
S_i(t) = -\sin(\omega_i t).
\eeq 
Consider $n$ such oscillators $i=1,2,\dots, n$, with fundamental frequencies $\omega_i= 2^{1-i}\pi\, \omega_0$. With time measured in units of $\omega_0^{-1}$, the period of oscillator $i$ is $2^i$, and so, by the arguments above, the output of oscillator $i$ at time $t_I=I(A)$, interpreted as 0 or 1 if less than or greater than zero, gives the $i^{\rm th}$ binary digit of the scaled analog input $I(A)$. That is, 
\beq
\label{eq_adc}
B_i = \Theta \left[S_i(t_I)\right],
\eeq
where $\Theta(x)$ is the Heaviside function, equal to 1 if $x>0$ and equal to zero otherwise. To this prescription we add the following caveat: if $I$ is an integer, we must observe the output of the oscillator at time $t_I=I(A)+\epsilon$, where $0<\epsilon<1$: the vertical black line in \f{fig_adc} indicates an observation time suitable for providing the binary representation of 14, namely 1110.

It is straightforward to verify that \eq{eq_adc}, with $i=0,1,2,3$, reproduces the entries $B_i$ of Table~\ref{tab1} when evaluated at integer $I$. But the formula also applies more generally, for non-integer $I$ and arbitrary bit number $n$. 

Each bit is calculated independently, and does not require digits to be carried between units. The calculation requires $n$ oscillators and $n$ comparators, and requires an observation time of up to $2^n-1$, in units of $\omega_0^{-1}$~\footnote{Note that oscillators made from RLC circuits can operate at frequencies $\omega_0$ of order GHz, comparable to the rate of a standard clock cycle on a CPU.}. If we consider only a single analog-to-digital conversion, then this approach is less efficient than a successive approximation register, which requires one comparator and operates in a time that scales linearly with the number of bits $n$ (flash analog-to-digital conversion operates in constant time, but requires $2^n-1$ comparators)\c{pelgrom2017}. 

However, the oscillatron design is intrinsically multifunctional in a temporal sense. When run for a time $t_I$, the outputs of the set of oscillators at all times less than $t_I$ provide the binary representation of all analog numbers less than $I$. Thus $n$ oscillators and $n$ comparators observed at $m$ different times would allow the analog-to-$n$-bit digital conversion of $m$ analog inputs, within a single run of the set of oscillators. The mean time required for the computation of each analog number then scales as $(2^n-1)/m$, while for a successive approximation register it is $mn$. For $m$ sufficiently large, the oscillator design is more efficient.

\new{In this section, for simplicity, we set the damping coefficient $\gamma$ to zero. The presence of damping will cause the magnitude of the oscillator to decay over time, limiting the time for which the oscillator's departure from zero can be measured, and so limiting the size of the analog number $I$ that can be represented. To counter this decay, energy can be input to the oscillator.}

\section{Training oscillatrons by gradient descent to perform multifunctional computation}
\label{grad}

\begin{figure*}
   \centering
   \includegraphics[width=\linewidth]{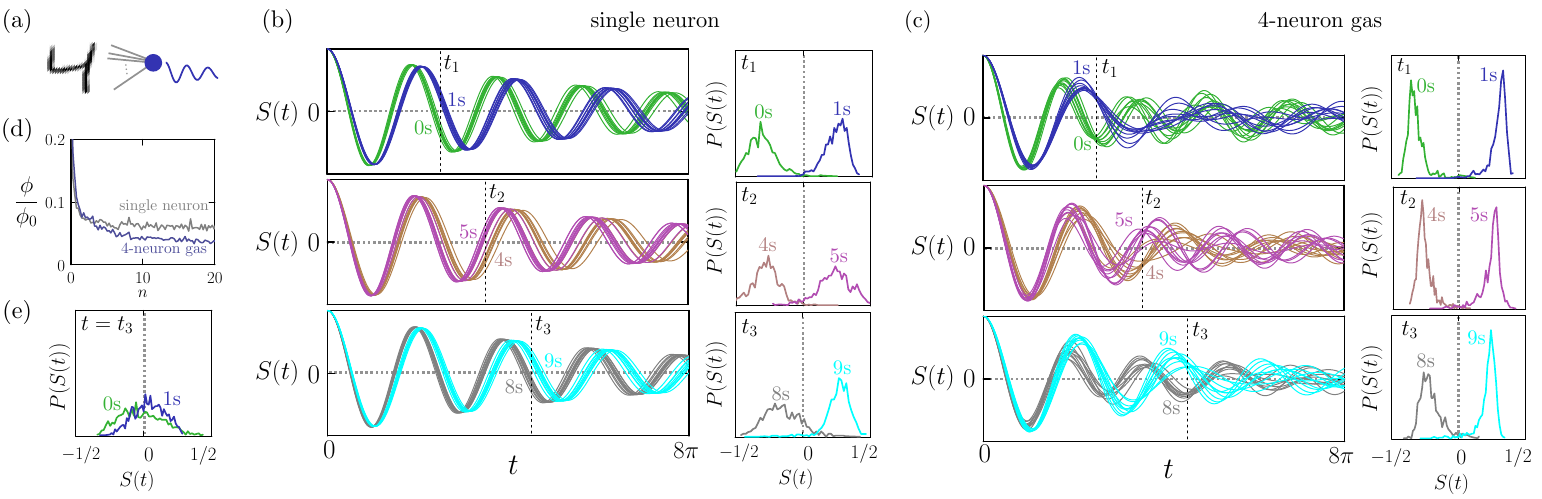} 
   \caption{Time-dependent multifunctional classification. (a) An oscillator neuron \eq{unit} with $N=28^2$ adjustable input weights is shown MNIST digits in classes $0,1,4,5,8$, and 9. Its weights are adjusted by gradient descent to distinguish 0s and 1s at time $t_1$; to distinguish 4s and 5s at time $t_2$; and to distinguish 8s and 9s at time $t_3$. (b) Time traces of the activity of the trained oscillator when shown 10 test-set digits (unseen during training) of each indicated class, and histograms of activity (using all digits of that class in the MNIST test set) at the indicated times. (c) As (b), but for a noninteracting gas of 4 oscillatrons, with the 4$N$ input weights adjusted by gradient descent to perform the same multi-time classification task. All histograms have a common vertical scale. (d) Loss \eq{phi}, scaled by $\phi_0=3 N_{\rm batch}$, as a function of training epoch $n$. (e) Activity histogram of the oscillator gas (when shown 0s and 1s) at a time at which it is not trained to discriminate 0s and 1s. \new{In panels (b), (c), and (e), the value zero is indicated horizontal and vertical grey dashed lines.} Oscillator parameters: $\omega_0=1, \gamma=1/10$.}
   \label{fig3}
\end{figure*}

In this section we show how an oscillator can be trained by gradient descent to achieve distinct tasks at distinct times. We use the oscillator \eq{unit} with intrinsic parameters $\omega_0=1$ and $\gamma=1/10$. We consider the MNIST data set\c{lecun1998gradient}, which consists of greyscale images of $70,000$ handwritten digits on a grid of $N=28 \times 28$ pixels, each digit belonging to one of ten classes $C \in [0,9]$; see \f{fig3}(a). We wish to construct a classifier that discriminates 0s (class $C=0$) from 1s (class $C=1$) at observation time $t_1$; discriminates 4s from 5s at time $t_2$; and discriminates 8s from 9s at time $t_3$. We choose observation times $t_1$, $t_2$, and $t_3$ such that $\Omega_0 t_k=- \arctan(2 \Omega_0/\gamma)+ (k+3) \pi$, where $\Omega_0^2 \equiv \omega_0^2-\gamma^2/4$, corresponding to three of the roots of the function \eq{sol} with $I=0$. 

A loss function suitable for this multi-time classification task is
\beq
\label{phi}
\phi= \sum_{k=1,2,3} \sum_{\alpha \in C_{4k},C_{4k+1}} (s_\alpha(t_k)-\bar{s}_\alpha)^2.
\eeq
Here $\alpha$ labels MNIST digits, and the inner sum runs over training-set digits in classes $C_{4k}$ and $C_{4k+1}$. We have defined $s_\alpha(t_k) \equiv {\rm e}^{\gamma t/2}S_{\alpha}(t_k)$; the exponential scaling cancels the exponential decay of the solution \eq{sol}, ensuring that each term in \eq{phi} is of equal importance. $S_{\alpha}(t_k)$ denotes the value of \eq{sol} at time $t_k$ when shown MNIST digit $\alpha$. The target activity in \eq{phi} is $\bar{s}_\alpha=\mp 1/2$, where the negative or positive sign is chosen if digit $\alpha$ is a member of an even-numbered class $(0,4,8)$ or an odd-numbered class $(1,5,9)$, respectively. The outer sum runs over the three observation times.

We consider $N$ connections between digit and neuron: when shown digit $\alpha$, the neuron input is $I_\alpha = \sum_{i=1}^N \theta_i I_{i,\alpha}$. Here $I_{i,\alpha}$ is the value of the $i^{\rm th}$ pixel of MNIST digit $\alpha$, and the $N$ quantities $\theta_i$ are adjustable weights. In what follows we use the notation $\Omega_\alpha^2 \equiv \omega_0^2-\gamma^2/4+I_\alpha$. 

To train an oscillator to solve this multi-time classification task we adjust its input weights by gradient descent on $\phi$. Weights are initially chosen randomly, $\theta_i \sim {\cal N}(0,10^{-4})$; we then iterate the equation 
\beq
\theta_i \to \theta_i - \alpha_0 \frac{\partial \phi}{\partial \theta_i}
\eeq 
for all $N$ weights $\theta_i$. Here $\alpha_0$ is the learning rate and 
\beq
\label{grad2}
\frac{\partial \phi}{\partial \theta_i}= \sum_{k=1,2,3} \sum_{\alpha \in C_{4k},C_{4k+1}} (s_\alpha(t_k)-\bar{s}_\alpha) g(t_k,\Omega_\alpha) I_{i,\alpha},
\eeq
where
\beq
\label{gee}
g(t,\Omega) \equiv \frac{\gamma t}{2 \Omega^2} \cos \Omega t -\frac{1}{\Omega}\left( t+ \frac{\gamma}{2 \Omega^2}\right) \sin \Omega t.
\eeq
We take the inner sum in \eq{grad2} over stochastically-chosen minibatches of $N_{\rm batch}=300$ digits, split equally between classes $C_{4k}$ and $C_{4k+1}$. We set the learning rate to $\alpha_0 = 10^{-3} (3 N_{\rm batch})^{-1}$~\footnote{It could happen during training that $\Omega_\alpha^2$ becomes negative for certain $\alpha$, indicating overdamping, in which case we must make the replacement $\Omega \to i \Omega$ in \eq{sol} and \eq{gee}. This did not happen in the simulations reported here.}. 

Results are shown in \f{fig3}(b). We plot time traces of oscillator activity when shown 10 test-set digits (unseen during training) in each class. Classes are indicated by color. The oscillator responds differently to each class, and can (imperfectly) distinguish the required classes at the required times: on the right of the panel we show histograms of oscillator activity, using all digits of the indicated class in the $10,000$-digit MNIST test set, at the designated observation times.

Note that training is done on the computer, the idea being that the trained weights can be implemented in hardware. \new{For instance, if currents are used to represent input pixel values, then resistors can be used to encode the weights}. This process is distinct to the process in which a physical system adapts to applied forces in order to learn to perform a task\c{stern2021supervised}.

As with standard neural units, collections of oscillatrons are more expressive than individual units. Multiple oscillators can have a total activity that is not periodic in time, even if they do not interact. In \f{fig3}(c) we show the analog of panel (b) for a collection of 4 noninteracting oscillator units (sometimes called a neural gas). Each has the same parameters $\omega_0$ and $\gamma$ as the single oscillator of panel (b), and each has $N$ input weights. We impose the loss function $\phi$, with $S_\alpha(t)$ replaced by $\frac{1}{4}\sum_{i=1}^4 S^{(i)}_\alpha(t)$. Here $S^{(i)}_\alpha(t)$ is the output of unit $i$ at time $t$ when shown MNIST digit $\alpha$. As before, we adjust the input weights by gradient descent on the loss. 

The time traces and histograms of \f{fig3}(c) show the trained oscillator gas to be more expressive than the trained individual oscillator, better distinguishing the required classes at the designated observation time. The loss values for the individual oscillator and the gas are shown in \f{fig3}(d). In panel (e) we show histograms of oscillator activity at a time $t_3$ for which the oscillator gas is trained to discriminate 8s and 9s but not 0s and 1s; there, it cannot distinguish 0s and 1s.

\section{Conclusions}
\label{conc} 

We have shown that the dynamics of an underdamped harmonic oscillator can perform multifunctional computation, with the same physical system able to solve distinct problems at distinct times within a single dynamical trajectory. This concept can be realized in hardware -- \eq{sol} is the output of the physical dynamics \eq{unit} -- and corresponds to a nonstandard form of oscillator computing: the latter usually focuses on the information contained within the phase of an oscillator, and seeks to identify the ground-state phases of coupled oscillators. Here we have considered the time-resolved amplitude of an oscillator whose inputs influence its frequency (or that is observed on a time that is a function of the input), which we interpret as a time-dependent neural unit or {\em oscillatron}, by comparison with perceptron. The activity of the oscillatron at fixed time is a nonmonotonic function of the input, and so it can solve nonlinearly-separable problems such as XOR. The activity of the oscillatron at fixed input is a nonmonotonic function of time, and so it is is multifunctional in a temporal sense, able to carry out distinct nonlinear computations at distinct times within the same dynamical trajectory. 

We have shown that oscillatrons can do all one-bit binary operations, function as all of the elementary logic gates, perform binary arithmetic, and convert an arbitrary number of analog values to a binary representation. We have also shown that oscillatrons can be trained by gradient descent to perform distinct classification tasks at distinct times. \new{To perform multifunctional computation we must observe oscillators at multiple times. Elements that can be used for timing (such as digital-to-time converters, phase-locked loops, or clocked latches) use power, and power expenditure would rise with the precision and number of time points needed\c{roberts2010brief,best2007pll}. The energy cost of timing is an important consideration to realize multifunctional computation in practice.}

A single oscillator can do nonlinear computation in a multifunctional way, and is a natural building block for neural networks and other devices. Neural networks can be built from interacting oscillators just as they are built from standard artificial neurons\c{atkins1991oscillatron,vodenicarevic2017neural,effenberger2022functional}. Oscillator neural nets designed for multifunctional computation would have considerable expressiveness. For instance, connecting $N$ oscillators $S_i(t)$ with springs or Ising-like potentials $U_{ij} = k_{ij} S_i(t) S_j(t)$ produces a neural network in which, for fixed input data, the activity of each oscillator is the superposition of $N$ normal modes, admitting neural waves for large $N$\c{frenchvibrationswaves,landaulifshitzmechanics}.  Neural oscillations and waves are common in the brain\c{kandel2000principles}. Inter-neuron couplings higher order than bilinear would allow for the possibility of chaotic behavior, resulting in stochastic behavior even in the absence of external noise; this would permit probabilistic and generative computation.

Computing with time-dependent functionality provides a way of carrying out multiple nonlinear computations within a single dynamical trajectory of a device, its natural time evolution giving us, in effect, multiple computations for the price of one. This idea could help reduce the number of parameters of a neural network or the number of circuit elements required to do specified computations.\\
  
{\em Acknowledgments---} I thank Isaac Tamblyn for comments on the paper. This work was done at the Molecular Foundry at Lawrence Berkeley National Laboratory, supported by the Office of Basic Energy Sciences of the U.S. Department of Energy under Contract No. DE-AC02--05CH11231.


%

\end{document}